# Indicating Asynchronous Multipliers

P. Balasubramanian, D.L. Maskell
School of Computer Science and Engineering
Nanyang Technological University
Singapore 639798
{balasubramanian, asdouglas}@ntu.edu.sg

N.E. Mastorakis
Department of Industrial Engineering
Technical University of Sofia
Sofia 1000, Bulgaria
mastor@tu-sofia.bg

*Abstract*—Multiplication is a basic arithmetic operation that is encountered in almost all general-purpose microprocessing and digital signal processing applications, and multiplication is physically realized using a multiplier. This paper discusses the physical implementation of indicating asynchronous multipliers, which are inherently elastic and are robust to timing, process, and parametric variations, and are modular. We consider the physical implementation of many weak-indication asynchronous multipliers using a 32/28-nm CMOS technology by adopting the array multiplier architecture. The multipliers are synthesized in a semi-custom ASIC-design style. The 4-phase return-to-zero (RTZ) and the 4-phase return-to-one (RTO) handshake protocols are considered for the data communication. The multipliers are realized using strong-indication or weak-indication full adders. Strong-indication 2-input AND function is used to generate the partial products in the case of both RTZ and RTO handshaking. The full adders considered are derived from different indicating asynchronous logic design methods. Among the multipliers considered, a weak-indication asynchronous multiplier utilizing the biased weak-indication full adder is found to be efficient in terms of the cycle time and the power-cycle time product with respect to both RTZ and RTO handshaking. Also, the 4-phase RTO handshake protocol is found to be preferable than the 4-phase RTZ handshake protocol for achieving enhanced optimizations in the design metrics.

*Keywords—multiplier, asynchronous circuits, indication, ASIC, standard cells, CMOS*

## I. INTRODUCTION

Addition is a basic arithmetic operation that forms the basis of other important arithmetic operations such as multiplication, division etc. Recently, in [1], different asynchronous implementations of the adder were discussed. This paper considers the robust asynchronous implementations of the multiplier since multiplication is also a common arithmetic operation that is encountered in almost all general-purpose microprocessing and digital signal processing applications [2]. References [3–9] discuss different transistor-level and gate-level designs of the asynchronous multipliers. However, most of these multipliers correspond to the bundled-data protocol, which has separate request and acknowledge wires besides the data bundle (i.e., data bus) and features a constant delay element with fixed delay assumed between the transmitter and the receiver. Due to the assumed delay for data transfer between the transmitter and the receiver, those multipliers are not robust when the presumed delay is exceeded, and therefore they are non-indicating.

In this work, we consider the robust class of asynchronous multipliers which are indicating. We consider the well-known array multiplier architecture for example, which corresponds to the shift-and-add multiplication approach. We physically implement many indicating asynchronous realizations of the 4×4 array multiplier, which utilize asynchronous components pertaining to different indicating asynchronous logic design methods. The resultant asynchronous array multipliers correspond to the weak-indication timing model.

The rest of this paper is organized into 4 sections. Section 2 gives a background into the design of indicating asynchronous circuits. Section 3 discusses different indicating asynchronous implementations of the 4×4 array multiplier by following a semi-custom ASIC design style. Section 4 presents the design metrics estimated for the multipliers after their physical realization using a 32/28-nm CMOS process technology. Also, the normalized power-cycle time product of the multipliers is provided. Finally, we draw some conclusions and state the scope for further work in Section 5.

## II. BASICS OF INDICATING ASYNCHRONOUS CIRCUIT DESIGN

### A. Data Encoding, Processing and Handshaking

The schematic of a typical indicating asynchronous circuit stage is shown in the middle of Fig. 1, which is correlated with the transmitter-receiver analogy at the top.

In Fig. 1, the current stage and the next stage registers are analogous to the transmitter and the receiver, and the indicating asynchronous circuit is sandwiched between the current stage and the next stage register banks. The register bank comprises a series of registers, with one register allotted for each of the rails of an encoded data input. Here, the register refers to a 2-input C-element. The C-element will output 1 or 0 if all its inputs are 1 or 0 respectively. If the inputs to a C-element are not identical, then the C-element would retain its existing steady-state. The circles with the marking 'C' represent the C-elements in the figures.

In Fig. 1, (A1, A0), (B1, B0) and (C1, C0) represent the delay-insensitive dual-rail encoded inputs of the single-rail inputs A, B and C respectively. According to dual-rail data encoding [10] and 4-phase RTZ handshaking [11], an input A is encoded as (A1, A0) where A = 1 is represented by A1 = 1 and A0 = 0, and A = 0 is represented by A0 = 1 and A1 = 0. Both these assignments are called *data*. The assignment A1 = A0 = 0 is called the *spacer*, and the assignment A1 = A0 = 1 is

This work is supported by the Academic Research Fund Tier-2 research award of Ministry of Education, Singapore under grant MOE2017-T2-1-002.

deemed illegal since the coding scheme should remain unordered [12] to maintain the delay-insensitivity.

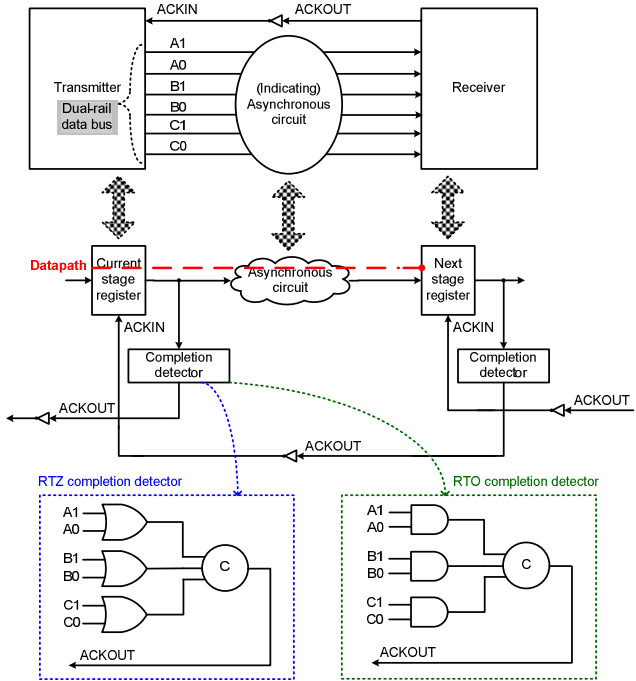

Fig. 1. A typical indicating asynchronous circuit stage. The RTZ and RTO completion detectors for the example dual-rail inputs are shown within the blue and green dotted boxes.

The application of input data to an indicating asynchronous circuit which conforms to the 4-phase RTZ handshake protocol follows this sequence: *data-spacer-data-spacer*, and so forth. It may be noted that the application of data is followed by the application of the spacer, which implies that there is an interim RTZ phase between the successive applications of input data. The RTZ phase ensures a proper data communication i.e., handshaking between the transmitter and the receiver. The RTZ handshaking process is governed by the following steps:

- Firstly, the dual-rail data bus specified by (A1, A0), (B1, B0) and (C1, C0) is a spacer, and therefore the acknowledgment input (ACKIN) is equal to binary 1. After the transmitter transmits a data, this would cause rising signal transitions i.e., binary 0 to 1 to occur on one of the dual rails of the entire dual-rail data bus
- Secondly, the receiver would receive the data sent and drive the acknowledgment output (ACKOUT) to 1
- Thirdly, the transmitter waits for ACKIN to become 0 and would then reset the dual-rail data bus, i.e., the dual-rail data bus becomes a spacer again
- Fourthly, after an unbounded but a finite and positive time duration, the receiver drives ACKOUT to 0 and subsequently ACKIN would assume 1. With this, a single data transaction is said to be complete, and the asynchronous circuit is permitted to start the next data transaction

According to the dual-rail data encoding and the 4-phase RTO handshaking [13], the input A is encoded as (A1, A0) but A = 1 is represented by A1 = 0 and A0 = 1, and A = 0 is represented by A0 = 0 and A1 = 1. Both these assignments are called *data*. The assignment A1 = A0 = 1 is called the *spacer*, and the assignment A1 = A0 = 0 is deemed illegal to maintain the delay-insensitivity.

The application of input data to an indicating asynchronous circuit conforming to the 4-phase RTO handshake protocol follows this sequence: *spacer-data-spacer-data*, and so forth. It may be noted that there is an interim RTO phase between the successive applications of input data. The RTO phase ensures a proper data communication i.e., handshaking between the transmitter and the receiver. The RTO handshaking process is governed by the following four steps:

- Firstly, the acknowledgment input (ACKIN) is equal to binary 1. After the transmitter transmits the spacer, this would cause rising signal transitions i.e., binary 0 to 1 on all the rails of the entire dual-rail data bus
- Secondly, the receiver would receive the spacer sent and drive the acknowledgment output (ACKOUT) to 1
- Thirdly, the transmitter waits for ACKIN to become 0 and would then transmit the data through the dual-rail data bus
- Fourthly, after an unbounded but a finite and positive time duration, the receiver drives ACKOUT to 0 and subsequently ACKIN would assume 1. With this, a single data transaction is said to be complete, and the asynchronous circuit is permitted to start the next data transaction

In an indicating asynchronous circuit, the time taken to process the data in the datapath highlighted by the red dashed lines in Fig. 1 is called *forward latency*, and the time taken to process the spacer is called *reverse latency*. Because there is an intermediate RTZ or RTO phase between the applications of two input data sequences, the *cycle time* is given by the sum of forward and reverse latencies. The cycle time of an indicating asynchronous circuit is the equivalent of the clock period of a synchronous circuit.

The gate-level detail of the example completion detectors corresponding to the 4-phase RTZ and RTO handshake protocols is shown at the bottom of Fig. 1, within the dotted blue and green boxes respectively. The completion detector indicates i.e., acknowledges the receipt of all the primary inputs given to an asynchronous circuit stage. In the case of the 4-phase RTZ handshaking, ACKOUT is provided by employing a 2-input OR gate to combine the respective dual rails of each encoded input, and then synchronizing the outputs of such 2-input OR gates using a C-element or a tree of C-elements. In the case of the 4-phase RTO handshaking, ACKOUT is provided by employing a 2-input AND gate to combine the respective dual rails of each encoded input, and then synchronizing the outputs of such 2-input AND gates using a C-element or a tree of C-elements. ACKIN is the Boolean complement of ACKOUT.

*B. Indicating Asynchronous Circuit Types*

Indicating asynchronous circuits are generally classified into two types as strong-indication and weak-indication [14]. The input-output timing correlation of strong-indication and weak-indication circuits is illustrated by a representative timing diagram, shown in Fig. 2. Strong-indication circuits [15] would wait to receive all the primary inputs (i.e., data or spacer) and would then process them to produce the required primary outputs (data or spacer). On the other hand, weak-indication circuits [16] can produce all but one of the primary outputs after receiving a subset of the primary inputs. Nevertheless, only after receiving the last primary input, they would produce the last primary output.

Both the strong- and weak-indication asynchronous circuits embed the isochronic forks assumption [17], which represents the weakest compromise to delay-insensitivity. Isochronic forks refer to the wires branching out from a node or junction, and the signal transitions whether they be rising or falling are presumed to occur concurrently on all the wire branches. But for the isochronic fork assumption, the practical realization of delay-insensitive circuits would not in fact be feasible [17]. It is reported in [18] that enforcing isochronicity is feasible even in the nanoelectronics regime, which is encouraging to note in the context of indicating asynchronous circuits.

A cascade of strong-indication sub-circuits may not result in a strong-indication circuit; rather, a weak-indication circuit may result. For example, if two strong-indication full adders are cascaded, the resultant would be a weak-indication 2-bit ripple carry adder (RCA). This is because if all the inputs to one of the full adders are provided, the corresponding sum and carry output bits of that full adder could be produced regardless of the non-availability of the inputs for the other full adder in the RCA. However, only after all the inputs to the other full adder are supplied, its corresponding sum and carry output bits would be produced. This scenario is characteristic of weak-indication, as discussed earlier.

Among the strong- and weak-indication circuits, the latter are preferable for practical implementation [19], and this is because of the strict timing restrictions inherent in the former. In general, for implementing arithmetic functions, the weak-indication type is preferable to the strong-indication type [20–22] and this is due to the following reasons: i) strong-indication arithmetic circuits tend to experience worst-case forward and reverse latencies for the application of data and spacer, and therefore the cycle time of strong-indication arithmetic circuits is always the maximum, ii) weak-indication arithmetic circuits may encounter data-dependent forward and reverse latencies or just a data-dependent forward latency and a constant reverse latency, and thus the cycle time of weak-indication arithmetic circuits is generally reduced compared to the strong-indication arithmetic circuits. However, for the weakly indicating asynchronous implementations of the array multiplier considered here, it is noted that their forward and reverse latencies would not be data-dependent or a constant; rather they correspond to the worst-case timing and so the cycle time also corresponds to the worst-case. Notwithstanding, the weak-indication array multipliers incorporating weak-indication full adders facilitate reductions in the cycle time, silicon area, and average power dissipation compared to the weak-indication array multipliers incorporating strong-indication full adders.

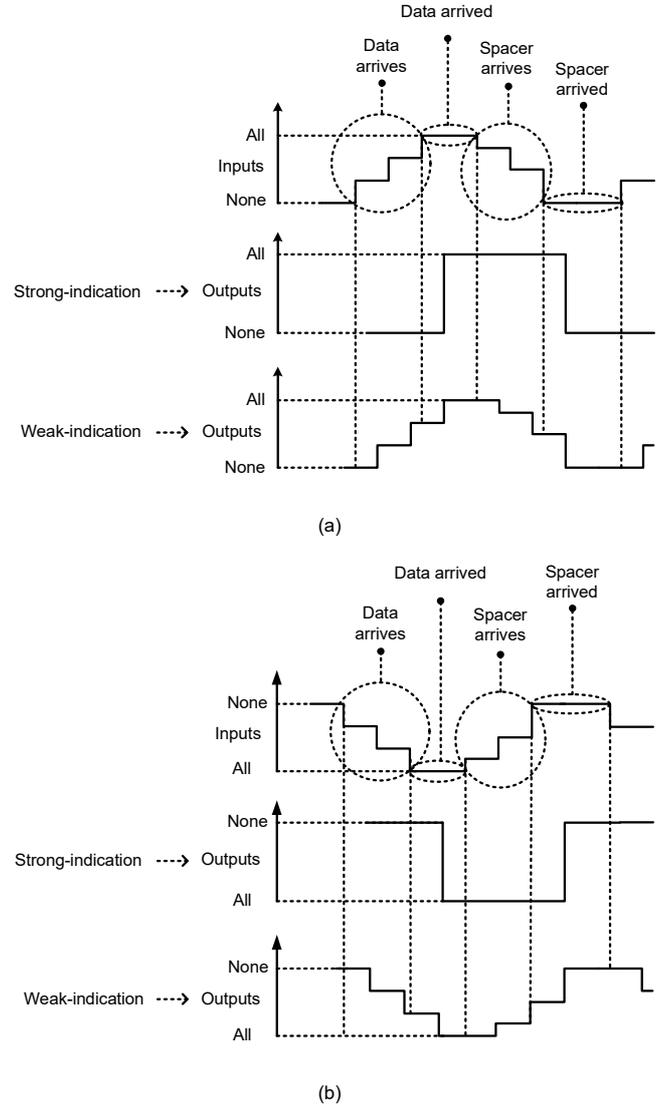

Fig. 2. Input-output timing relation of strong-indication and weak-indication circuits corresponding to: (a) RTZ handshaking, and (b) RTO handshaking.

III. INDICATING ASYNCHRONOUS ARRAY MULTIPLIERS

The 4×4 array multiplier structure is portrayed by Fig. 3. Here, (A3, A2, A1, A0) and (B3, B2, B1, B0) represent the inputs of the multiplier which are dual-rail encoded. (A3, B3) and (A0, B0) represent the most significant and the least significant input bit-pairs respectively. P7 to P0 represent the product bits, which are also dual-rail encoded, with P7 being the most significant product bit and P0 being the least significant product bit.

Fourteen indicating asynchronous array multipliers were physically realized with seven multipliers corresponding to the RTZ handshake protocol and the same seven multipliers corresponding to the RTO handshake protocol. The intent is to determine which indicating asynchronous logic components would be more optimum to realize the array multiplier. This observation may also be useful to determine which indicating asynchronous logic components would be better suited for the optimum realization of indicating asynchronous multipliers corresponding to the other multiplier architectures. Further, it is of interest to ascertain whether the RTZ or the RTO handshake protocol would help to better optimize the design metrics.

The 4×4 array multiplier requires sixteen 2-input AND functions to generate the partial products and twelve full adders. Of these, the carry input of four full adders are set to 0 in the case of RTZ handshaking and set to 1 in the case of RTO handshaking. The inputs to the full adders in the multiplier array shown in Fig. 3 represent the partial products. These partial products are generated through the 2-input AND function. The strong-indication realization of the 2-input AND function corresponding to RTZ and RTO handshake protocols are portrayed by Figs. 4(a) and 4(b) respectively. In Fig. 4, C1 to C4 represent the 2-input C-elements. (X1, X0) and (Y1, Y0) are the inputs of the 2-input AND function and (Z1, Z0) is its output.

- Two indicating asynchronous array multipliers which incorporate strong-indication full adders based on [26], corresponding to RTZ and RTO handshake protocols
- Two indicating asynchronous array multipliers which incorporate strong-indication full adders based on [27]; corresponding to RTZ and RTO handshake protocols
- Two indicating asynchronous array multipliers which incorporate strong-indication full adders based on [28]; corresponding to RTZ and RTO handshake protocols
- Two indicating asynchronous array multipliers which incorporate weak-indication full adders based on [27]; corresponding to RTZ and RTO handshake protocols
- Two indicating asynchronous array multipliers which incorporate weak-indication full adders based on [29]; corresponding to RTZ and RTO handshake protocols
- Two indicating asynchronous array multipliers which incorporate weak-indication full adders based on [30]; corresponding to RTZ and RTO handshake protocols
- Two indicating asynchronous array multipliers which incorporate weak-indication full adders based on [31]; corresponding to RTZ and RTO handshake protocols

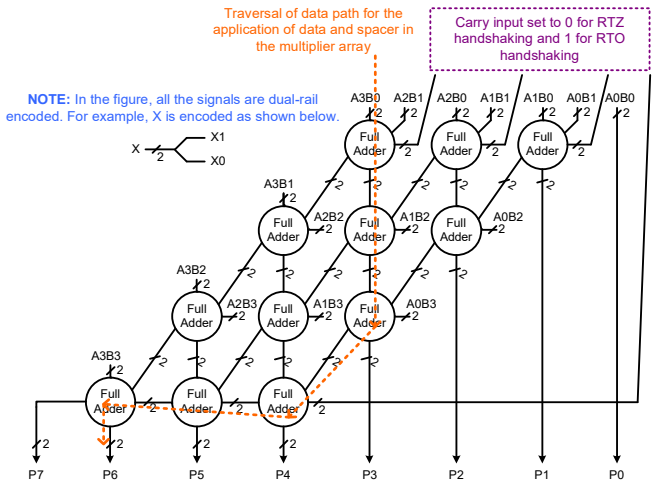

Fig. 3. 4×4 array multiplier schematic. The partial products, primary inputs, intermediate outputs, and primary outputs shown are all dual-rail encoded.

References [1] [23] [41] provide practical examples for the transformation of an asynchronous logic corresponding to the RTZ protocol into that that corresponds to the RTO protocol and vice-versa. The rules for the logical transformation between the RTZ and RTO handshake protocols are given in [25] along with the proofs. Note that a weak-indication 2-input AND function cannot be physically realized since it has one dual-rail primary output. A weak-indication design requires at least a pair of dual-rail primary outputs to satisfy the weak-indication timing constraints [14].

The indicating asynchronous full adders derived from different logic design methods [26–31] are used to realize the asynchronous array multipliers, by substituting the full adders in the respective places as highlighted in the architecture shown in Fig. 3. The fourteen asynchronous array multipliers are implemented as follows:

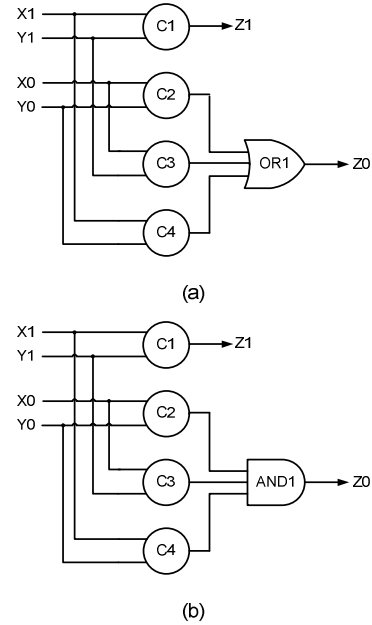

Fig. 4. Strongly indicating realization of the 2-input AND function corresponding to: (a) RTZ handshaking, and (b) RTO handshaking.

IV. RESULTS AND DISCUSSION

Fourteen indicating asynchronous array multipliers were physically realized using the gates of a 32/28-nm bulk CMOS standard digital cell library [32], and all the array multipliers correspond to weak-indication. The 2-input C-element does not form a part of the cell library and so it was custom-realized using the AO222 cell with feedback to implement the various array multipliers. Delay-insensitivity was carefully considered while decomposing the logic [28] [33] to avoid the possibility of gate orphan(s). A gate orphan is an unacknowledged signal transition on a gate output. Gate orphans are problematic as they might affect the robustness of an indicating asynchronous circuit and so they are better avoided [34]. For a detailed explanation of gate orphans, the interested reader is referred to [35–37]. Wire orphan, which refers to the unacknowledged transition on a data wire, is however nullified by imposing the isochronic fork assumption [38].

A typical case PVT specification of the high $V_t$ digital cell library viz. 1.05V and 25°C was used to perform the simulations. The design metrics such as cycle time, area, and average power dissipation estimated are given in Table I.

TABLE I. DESIGN METRICS OF INDICATING ASYNCHRONOUS MULTIPLIERS CORRESPONDING TO RTZ AND RTO HANDSHAKING, ESTIMATED USING A 32/28-NM CMOS PROCESS TECHNOLOGY

| Multiplier Reference | Cycle Time (ns) | Area (µm²) | Power (µW) | Normalized PCTP |
|---|---|---|---|---|
| *Corresponding to 4-phase RTZ handshake protocol* | | | | |
| [26] | 7.26 | 1015.30 | 1245 | 1 |
| [27][1] | 5.42 | 1006.16 | 1228 | 0.736 |
| [28] | 5.32 | 926.86 | 1207 | 0.710 |
| [27][2] | 5.20 | 975.66 | 1222 | 0.703 |
| [29] | 5.18 | 823.17 | 1216 | 0.697 |
| [30] | 3.90 | 853.67 | 1222 | 0.527 |
| [31] | 4.48 | 835.37 | 1217 | 0.603 |
| *Corresponding to 4-phase RTO handshake protocol* | | | | |
| [26] | 7.08 | 1015.31 | 1240 | 1 |
| [27][1] | 5.16 | 957.36 | 1211 | 0.712 |
| [28] | 5.24 | 926.86 | 1206 | 0.720 |
| [27][2] | 5.02 | 951.26 | 1211 | 0.692 |
| [29] | 5.12 | 823.17 | 1212 | 0.707 |
| [30] | 3.70 | 853.67 | 1217 | 0.513 |
| [31] | 4.38 | 835.37 | 1213 | 0.605 |

[1] Utilizes strong-indication full adder; [2] Utilizes weak-indication full adder

As mentioned earlier, the cycle time of an indicating asynchronous circuit is synonymous with the clock period of a synchronous circuit. Given that the cycle time is the sum of forward and reverse latencies, the forward latency is like the critical path delay which can be directly estimated. The estimation of reverse latency is non-trivial since it is the time taken to process the spacer. The reverse latency cannot be directly estimated using a commercial static timing analyzer, and so the reverse latency was estimated manually based on the timing information derived from the gate-level simulations. For the indicating asynchronous multipliers mentioned in Table I, their forward and reverse latencies are equal, and the longest datapath traversed for the application of data or spacer is the same, which is highlighted by the dotted orange line in Fig. 3.

Since power and cycle time are desirable to be less, the power-cycle time product (PCTP) is also desirable to be less. The PCTP serves as a qualitative low power parameter for an indicating asynchronous design, which is analogous to the power-delay product of a synchronous design. The PCTP of the indicating asynchronous multipliers given in Table I are calculated and normalized. To perform the normalization, the highest value of the PCTP corresponding to a multiplier was considered as the baseline, and this value was used to divide the actual PCTPs of all the multipliers. This procedure was followed for the multipliers corresponding to the RTZ and RTO handshake protocols separately. Thus, the least value of the PCTP is representative of the best design in Table I with respect to RTZ and RTO handshaking.

It is seen from Table I that the average power dissipation does not vary significantly across the different multipliers, and this is because all the indicating asynchronous array multipliers satisfy the monotonic cover constraint (MCC) [11]. The MCC basically refers to the activation of a unique signal path from a primary input to a primary output after the application of an input data. The MCC enables ensuring the proper indication of signal transitions throughout an entire asynchronous circuit from the first up to the last logic level. This is because the signal transitions, whether they be rising or falling, should occur monotonically throughout the entire circuit [34], and the MCC ensures this. The MCC arises from the adoption of a logic expression format which is composed of disjoint or orthogonal terms [39] to describe the primary outputs. For example, in a disjoint sum-of-products expression, the logical conjunction of any two product terms would yield 0 [40]. Hence, only one term would become activated in a disjoint logic expression after the application of data.

It can be seen from Table I that the weak-indication array multiplier incorporating the biased weak-indication full adder of [30] and the strong-indication 2-input AND function to generate the partial products enables reduced cycle time and PCTP compared to the rest with respect to RTZ and RTO handshaking. Compared to the weak-indication array multiplier embedding the weak-indication full adder of [31], the weak-indication array multiplier embedding the weak-indication full adder of [30] reports corresponding reductions in cycle time and PCTP by 12.9% and 12.6% for RTZ handshaking, and by 15.5% and 15.2% for RTO handshaking respectively.

V. CONCLUSION AND SCOPE FOR FURTHER WORK

This paper has discussed the physical implementation of robust indicating asynchronous array multipliers based on indicating asynchronous logic design methods. The multipliers were realized based on 4-phase RTZ and RTO handshake protocols, and they correspond to the weak-indication. It is noted that the array multiplier incorporating the weak-indication full adder of [30] enables enhanced optimizations in the design metrics compared to the other indicating asynchronous array multipliers. As the size of the multiplication is increased, we hypothesize that the array multiplier utilizing the weak-indication full adder of [31] might be competitive to that utilizing the weak-indication full adder of [30]. However, higher bit-width multiplications should have to be considered to unravel the reality. Nevertheless, both [30] and [31] present full adder designs which incorporate redundant logic, and it was shown in [43] that logic redundancy could help to significantly reduce the latencies and the cycle time at almost no increase in the area or average power dissipation.

The construction of indicating asynchronous array multipliers given in Table I is quite straightforward since the full adders based on the corresponding design methods [26–31] can be used to substitute the respective function blocks as shown in Fig. 3, corresponding to RTZ or RTO handshaking. However, the constructions of early output quasi-delay-insensitive asynchronous array multipliers using early output full adders of say, [22] and [42] may not be straightforward. This is due to the likelihood of the problem of gate orphans. To overcome the gate orphan problem in the realization of the early output asynchronous array multiplier, the provision of internal completion detectors as used in [44] may become a necessity to ensure a proper indication of rising and falling signal transitions at the intermediate outputs. Moreover, the outputs of all the internal completion detectors would have to be synchronized with at least one dual-rail product bit of the array multiplier using a tree of C-elements. This would enable the provision of proper acknowledgment for the receipt of data or spacer throughout an indicating asynchronous circuit starting from the first logic level up to the last logic level.

Although the early output logic could simplify the physical realization of the full adder blocks thereby suggesting potential savings in the design metrics, the additional introduction of internal completion detectors to ensure delay-insensitivity may partially or fully offset the reductions in the design metrics achieved for the early output logic type. However, this should be investigated. Hence, the design and implementation of early output quasi-delay-insensitive asynchronous array multipliers and their comparison with the indicating asynchronous multipliers in terms of the design metrics is necessary, which points to a scope for further work.